\documentclass[twocolumn,aps,showpacs,preprintnumbers,amsmath,amssymb,superscriptaddress]{revtex4-1}

\usepackage[pdftex]{graphicx}
\usepackage{amsmath}% bold math
\usepackage{color}
\usepackage{ulem}
\usepackage{float}
\usepackage{color}
\begin{document}
\title{Thermoelectric spin accumulation and long-time spin precession in a non-collinear quantum dot spin valve}
\bigskip
\author{Bhaskaran Muralidharan}
%author[2]{Milena Grifoni}
%\affiliation{Institut f\"ur Theoretische Physik, Universit\"at Regensburg, Regensburg D-93040, Germany}
\affiliation{Department of Electrical Engineering and Center of Excellence in Nanoelectronics, Indian Institute of Technology Bombay, Powai, Mumbai-400076, India}
\affiliation{Institut f\"ur Theoretische Physik, Universit\"at Regensburg, Regensburg D-93040, Germany}
\author{Milena Grifoni}
%author[2]{Milena Grifoni}
\affiliation{Institut f\"ur Theoretische Physik, Universit\"at Regensburg, Regensburg D-93040, Germany}
\date{\today}
\medskip
\widetext
\begin{abstract}
We explore thermoelectric spin transport and spin dependent phenomena in a non-collinear quantum dot spin valve set 
up. Using this set up, we demonstrate the possibility of a thermoelectric excitation of single spin 
dynamics inside the quantum dot. Many-body exchange fields generated on the single spins in this set up manifest as effective 
magnetic fields acting on the net spin accumulation in the quantum dot. We first 
identify generic conditions by which a zero bias spin accumulation in the dot may be 
created in the thermoelectric regime. 
The resulting spin accumulation is then shown to be subject to a field-like spin torque due to the effective 
magnetic field associated with either contact. This spin torque that is 
generated may yield long-time precession 
effects due to the prevailing blockade conditions. The implications of these phenomena in connection with single spin manipulation and pure spin current generation are then discussed.
\end{abstract}
\maketitle
\section{Introduction}  Achieving the control of individual spins \cite{Leo_review} or collective spin degrees of freedom \cite{Bader} forms an important frontier of spintronics. In the collective case, the manipulation of magnetization dynamics \cite{Ralph_1,Ralph_2} in magnetic nanostructures via spin transfer torques \cite{Ralph_Stiles_2007} or magnetic domain wall dynamics \cite{Bader,Parkin} has been a topic of much attention. At the same time, quantum dots provide an ideal platform for realizing individual spin manipulation and control \cite{Leo_review}. Control of magnetization dynamics forms the basis of a wide range of applications from microwave oscillators to magnetic storage \cite{Ralph_1,Ralph_2}, while that of individual spins is an important paradigm towards spin based quantum computation \cite{Loss}. 
\\ \indent The control and manipulation of individual spins in quantum dots \cite{SB_1,SB_2,SB_3} has become possible owing to the ability to lock the number of electrons, as well as their individual spins. While the electron number can be controlled by a gate voltage due to Coulomb blockade, their net spin accumulation may be controlled via spin blockade \cite{Leo_review,Tarucha_1}. Spin blockade is a condition when an electron current flow under non-equilibrium conditions is forbidden due to the interplay between Pauli exclusion principle and Coulomb interaction. The net spin accumulation via the spin blockade mechanism may also be fine tuned via the use of a gate electrode. In a typical control experiment, a gate electrode pulses the system in and out of spin blockade, thus permitting spin manipulation when the electron current flow is forbidden, and read out when the current flow is permitted \cite{SB_1,SB_2,SB_3}. 
\\ \indent Spin blockade and spin manipulation in the aforementioned works were discussed under a voltage bias. In recent times, there has been a lot of activity in the area of spin based thermoelectrics or spin caloritronics \cite{Bauer_2011} and hence it is timely to investigate spin transport under the application of a temperature gradient \cite{Jansen}. Specifically, spin dependent thermoelectric effects in quantum dots have also been theoretically investigated in a few recent works \cite{Barnas_1,DiVentra_spin,Wang,Swirkowicz,Slovenia,Sothmann}. In this paper we explore the possibility of spin manipulation by creating a non-equilibrium spin accumulation in the thermoelectric regime. While in the pioneering spin manipulation experiments \cite{Leo_review,SB_1,SB_2,SB_3} spin blockade occurs due to a blocking triplet state \cite{Basky_Datta} in a detuned double quantum dot set up with unpolarized contacts, we focus on creating the spin accumulation via a different spin blockade mechanism in a non-collinear quantum dot spin valve described extensively in some earlier works \cite{Koenig_1,Koenig_2}. In our set up, unlike in the double quantum dot case, the spin blockade results from the spin selection and filtering between spins in the quantum dot and the ferromagnetic degrees of freedom of the contacts whose magnetization directions in general, may be non-collinear. Also, in our set up, many-body exchange fields are generated from an interplay between the Coulomb interaction in the dot or metallic island and the ferromagnetic degree of freedom in the contacts \cite{Koenig_1,Koenig_2,Grifoni_Bauer}. The effective magnetic field thereby creates a {\it{field-like}} term in the description of the spin dynamics inside the dot. This field like term is reminiscent of spin torque in magnetic structures \cite{Ralph_Stiles_2007} and is responsible for the precessional spin dynamics inside the dot. In addition to the precessional term, one has terms arising from the spin polarized current injection, as well as relaxation due to single-electron tunneling processes between either contact and the dot. 
\\ \indent We show here that the precessional term that arises out of the above mentioned field-like spin torque may be created under a pure thermal gradient in the absence of a bias. The crucial aspect is that the non-equilibrium spin accumulation is induced as a result of a spin blockade mechanism, to be discussed, in the regime where double occupancy is suppressed due to Coulomb interaction. As a result of long dwell times in the dot due to the blockade, the charge and spin relaxation components of the spin dynamics are suppressed, thus yielding a long time precession. 
\\ \indent This field like spin torque itself translates to a net spin angular momentum transfer rate or a spin current between the contacts and the dot \cite{Koenig_3}. The traditional viewpoint of a spin current is that of a spin polarized current resulting from the transport of spin polarized electrons. The precessional terms also imply a net angular momentum transfer rate mediated by exchange interaction, and may, in general, also be affiliated with spin currents \cite{Koenig_3}. Such spin currents resulting from a spin precession may possibly be detected via optical or electrical means as demonstrated in some recent pioneering experiments \cite{C_Back,Brataas_3}. Earlier works on spin dependent thermoelectrics in quantum dots  primarily focused either on the linear response thermoelectric regime \cite{Barnas_1,Slovenia}, or on the generation of pure spin currents using non-magnetic quantum dots in the presence of a magnetic field \cite{DiVentra_spin}, or magnetic quantum dots \cite{Wang} with collinearly polarized contacts, or novel effects that arise due to the coupling with magnons \cite{Sothmann}. But these works, however, do not feature the effects related to spin precession to be discussed here.
\\ \indent The paper is organized as follows. The following section will describe the necessary formulation briefly, and will cover the important aspects of the physics of angular momentum transfer in relation to its coverage in this paper. We then discuss the important results and their implications in section III. Section IV concludes the paper. 
\section{Set up and formulation} 
\begin{figure}
	\centering
		\includegraphics[width=3.3in,height=2.8in]{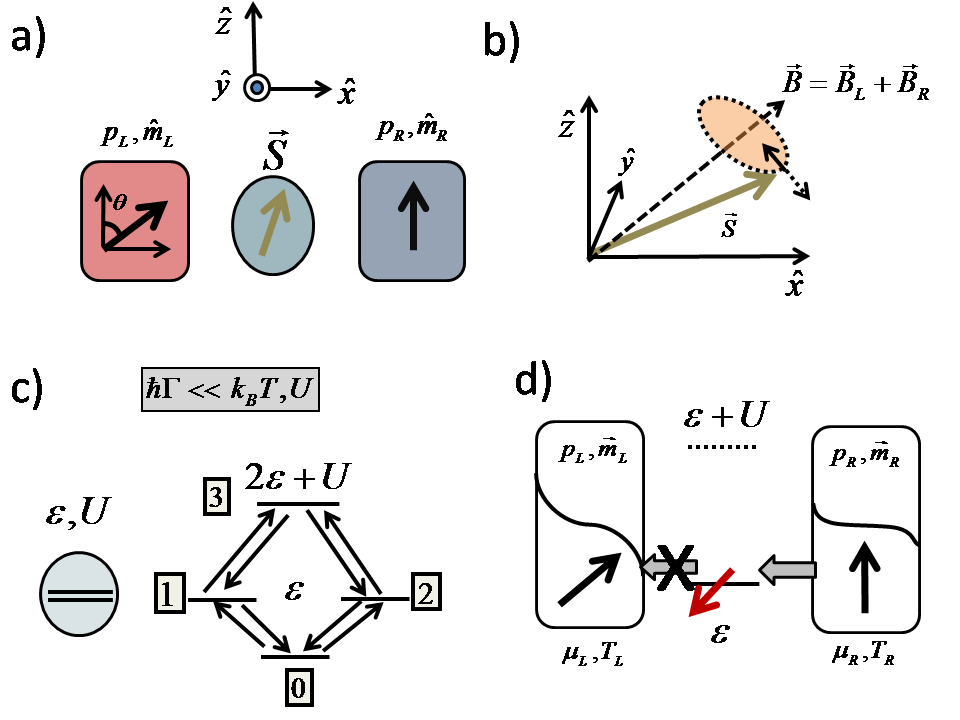}
		\caption{(Color online) Non-collinear quantum dot spin valve transport set up. a) The set up consists of a quantum dot weakly coupled to ferromagnetic contacts $\alpha=L,R$, each with a pinned 
magnetization axis $\hat{m}_{\alpha}$ oriented along the majority spin and a degree of polarization $p_{\alpha}$. The contact $L(R)$ acts as the collector (injector) in the forward (reverse) bias direction. The common coordinate axis is chosen to be oriented with respect to that of the quantum dot, with the $\hat{x}$ axis being pointing in the (longitudinal) transport dimension. The angle between the contact magnetizations is $\theta$. The set up may be spin blockaded for a certain range of bias and for certain values of $\theta$. b) Spin dynamics comprise of spin precession around the net direction of the effective exchange field $\vec{B}_L+\vec{B}_R$ and relaxation introduced via charge tunneling 
to and from the contacts. c) Transport through the 
single level quantum dot in the sequential tunneling regime is modeled via density matrix rate equations which may be viewed as transitions between many-electron states labeled $0$ through $3$. d) Example of a transport set up that displays spin blockade when the average electron number inside the quantum dot is unity. Spin blockade results in an accumulation of spins antiparallel to the collector contact spin polarization.}  
	\label{fig:sp_fig1}
\end{figure}

In the schematic of the quantum dot spin valve set up shown in Fig.~\ref{fig:sp_fig1}(a), the quantum dot is weakly coupled to two non-collinearly polarized ferromagnetic contacts labeled $\alpha=L,R$, each with a degree of polarization $p_{\alpha}$, an electrochemical potential $\mu_{\alpha}$, and a temperature $T_{\alpha}$. The contact $L(R)$ acts as the collector (injector) in the forward (reverse) bias direction. Second order transport theory across quantum dots weakly coupled to ferromagnetic contacts predicts that the interplay between the strong Coulomb repulsion in the dot and the spin polarization of the itinerant electrons to and from the ferromagnetically pinned contacts results in a many body exchange field like term \cite{Brouw,Koenig_1,Koenig_2} that drives the precessional dynamics inside the dot. The non-equilibrium spin dynamics of the quantum dot spin accumulation $\vec{S}$ is composed of spin injection, relaxation and precession terms \cite{Koenig_2}, as shown in Fig.~\ref{fig:sp_fig1}(b).
\subsection{Model}
The theoretical description of transport in our set up begins by defining the overall Hamiltonian $\hat{H}$ 
which is usually written as $\hat{H}=\hat{H}_D + \hat{H}_{C} + \hat{H}_{T}$, where $\hat{H}_D, \hat{H}_C$ and $\hat{H}_{T}$ represent the dot, 
reservoir and reservoir-dot coupling Hamiltonians, respectively. In this paper, the quantum dot is modeled as a single orbital Anderson impurity described by the one-site Hubbard Hamiltonian:
\begin{equation}
\hat{H}_D=\sum_{\sigma} \epsilon_{\sigma} \hat{n}_{\sigma} + U \hat{n}_{\uparrow} \hat{n}_{\downarrow},
\label{Ham_def}
\end{equation}
where $\epsilon_{\sigma}$ represents the orbital energy, $\hat{n}_{\sigma} = \hat{d}_{\sigma}^{\dagger} \hat{d}_{\sigma}$ is the occupation number operator of an electron with 
spin $\sigma = \uparrow$, or $\sigma = \downarrow$, and $U$ is the Coulomb interaction energy between electrons of opposite spins occupying the same orbital. The exact diagonalization 
of the dot Hamiltonian then results in four Fock-space energy levels labeled by their total energies 
$0$, $\epsilon_{\uparrow}$, $\epsilon_{\downarrow}$, and $\epsilon_{\uparrow}+\epsilon_{\downarrow} +U$. 
In this paper, we consider only a spin-degenerate level such that $\epsilon=\epsilon_{\uparrow}=\epsilon_{\downarrow}$. 
The contact Hamiltonian is given by 
$\hat{H}_{C} = \sum_{\alpha=L,R}\sum_{k \sigma_{\alpha}} \epsilon_{\alpha k \sigma_{\alpha}} \hat{n}_{\alpha k \sigma_{\alpha}}$, 
where $\alpha$ labels the left/right reservoir ($L$ or $R$ in our case) and the summation is taken over the single particle states labeled 
$\{k \sigma_{\alpha}\}$, and $\sigma_{\alpha}=\pm$ denotes the majority and minority spin orientation in the contacts. 
The tunneling Hamiltonian that represents the dot-contact coupling may in general be written as:
\begin{equation} 
\hat{H}_{T}=\sum_{\alpha k \sigma_{\alpha}} \left ( t_{\alpha} \hat{c}^{\dagger}_{\alpha k \sigma_{\alpha}} \hat{d}_{\sigma_{\alpha}} + 
{{t}^{\ast}_{\alpha}} \hat{d}^{\dagger}_{\sigma_{\alpha}} 
\hat{c}_{\alpha k \sigma_{\alpha}} \right ) %\nonumber\\
%&=& \sum_{\alpha k \sigma' \sigma} \hat{h}_{T\alpha k \sigma' \sigma}, 
\label{eq:Tunn_Ham}
\end{equation}
where $(\hat{c}^{\dagger},\hat{c})$ and $(\hat{d}^{\dagger},\hat{d})$ are the creation/annihilation operators of the reservoir states labeled $\{k \sigma_{\alpha} \}$ and of the quantum dot one particle states respectively, and $t_{\alpha}$ denotes the tunneling matrix element. Note that, in general, the direction of majority and minority spins $\sigma_{\alpha}=\pm$ in either contact and of the spin orientation $\sigma$ $=$ $\uparrow, \downarrow$ in the dot may be non-collinear. If the $\hat{z}$ axis of the spin polarization in contact $\alpha$ makes an angle $( \theta_{\alpha},\phi_{\alpha} )$ with the $\hat{z}$ axis of the dot, one can rewrite the tunneling Hamiltonian in Eq. (\ref{eq:Tunn_Ham}) as \cite{Koenig_1}:
 \begin{eqnarray} 
\hat{H}_{T}&=&\sum_{\alpha k  } \left ( t_{\alpha} \hat{c}^{\dagger}_{\alpha k +} (C_{\alpha} \hat{d}_{\uparrow} + S_{\alpha} \hat{d}_{\downarrow}) \right )\nonumber\\
&+& \sum_{\alpha k } \left ( t_{\alpha} \hat{c}^{\dagger}_{\alpha k -} (-S^{\ast}_{\alpha} \hat{d}_{\uparrow} + C^{\ast}_{\alpha} \hat{d}_{\downarrow}) \right ) + h.c.,
\label{eq:Tunn_Ham_2}
\end{eqnarray}
where $C_{\alpha}=\cos(\theta_{\alpha}/2)e^{i\phi_{\alpha}/2}, S_{\alpha}=\sin(\theta_{\alpha}/2)e^{-i\phi_{\alpha}/2}$, and $h.c.$ stands for the hermitian conjugate. In this work, without loss of generality, we let $\phi=0$ so that the orientations of the magnetization directions of the two contacts are in the $\hat{x}-\hat{z}$ plane.
We can then define the tunneling rate for each spin $\sigma_{\alpha}$ associated with contact $\alpha$ as $\Gamma_{\alpha \sigma_{\alpha}}= \frac{2 \pi}{\hbar} \sum_{k}|t_{\alpha}|^2 \delta(E-\epsilon_{\alpha k \sigma_{\alpha}})=\frac{2 \pi}{\hbar} |t_{\alpha}|^2 D_{\alpha \sigma_{\alpha}}$, where $D_{\alpha \sigma_{\alpha}}$represent the density of states (assumed constant in our case) of the majority and minority spins of the contact. We can then define a degree of polarization associated with either contact as $p_{\alpha}=(\Gamma_{\alpha +}-\Gamma_{\alpha -})/(\Gamma_{\alpha +}+\Gamma_{\alpha -})=(\Gamma_{\alpha +}-\Gamma_{\alpha -})/\Gamma_{\alpha}$.
\subsection{Spin accumulation and spin currents}
The calculation of the non-equilibrium spin accumulation $\vec{S}$ and of its dynamics follows from the evaluation of the reduced density matrix of the dot using the density matrix formulation discussed extensively in \cite{Koenig_2}. For this, one starts with the time evolution of the composite (dot + contacts) density matrix $\hat{\rho}(t)$ which is given by the Liouville equation. The reduced density matrix $\hat{\rho}_{red}(t)$ of the dot is then obtained by performing a trace exclusively over the reservoir space. An expansion of the Liouville equation up to the second order in the tunneling Hamiltonian in the limit of weak contact coupling $(\hbar \Gamma \ll k_BT)$ leads to the density matrix master equation for the reduced density matrix of the system \cite{Koenig_1,Brouw,Timm,Milena_noncoll}. In this paper, we consider the regime of sequential tunneling $\hbar \Gamma_{\alpha} << k_BT$, $U$ such that this description based on a second order perturbation in the tunnelling matrix element $t_{\alpha}$ will suffice. In this regime, transport as described via density matrix rate equations for the reduced density matrix of the dot \cite{Brouw} may be viewed as transitions between Fock space states of the dot as depicted in Fig.~\ref{fig:sp_fig1}(c). We consider steady state transport in all our calculations and hence consider the steady state solution $\rho_{ij}$ of the reduced density matrix of the dot. The diagonal terms $\rho_{ii}$ of this density matrix represent the probability of occupation of each many electron state $i $ labeled $0$ through $3$. The average spin of the dot along its $\hat{z}$ direction is given by $S_z=\frac{\hbar}{2}\left ( \frac{\rho_{11}-\rho_{22}}{2} \right )$. 
The off-diagonal terms $\rho_{12}$, $\rho_{21}$ relate to the average spin in the quantum dot along the remaining 
two axes such that $S_x=\frac{\hbar}{2}\left ( \frac{\rho_{12}+\rho_{21}}{2} \right )$, $S_y=i\frac{\hbar}{2}\left ( \frac{\rho_{12}-\rho_{21}}{2} \right )$. The spin dynamics associated with the non-equilibrium spin accumulation $\vec{S}$ are then described by \cite{Koenig_2,Milena_noncoll}:
\begin{eqnarray}
\frac{2q}{\hbar}\frac{d\vec{S}}{dt}&=&\sum_{\alpha} \left [ J^q_{\alpha}p_{\alpha}\hat{m}_{\alpha} - \frac{2q}{\hbar} \left ( \frac{\vec{S}-p^2_{\alpha}(\hat{m}_{\alpha} \cdot \vec{S})\hat{m}_{\alpha}}{\tau_{r,\alpha}} \right ) \right ] \nonumber\\
&-&  \frac{2q}{\hbar} \sum_{\alpha} \vec{S} \times \vec{B}_{\alpha},
\label{eq:spin_dyn}
\end{eqnarray}
with $J^q_{\alpha}$ being the terminal charge current, $-q$ being the magnitude of the electron charge, $p_{\alpha}$ being the degree of polarization of each contact, and $1/\tau_{r,\alpha}=\Gamma_{\alpha} \left (1-f_{\alpha}(\epsilon)+f_{\alpha}(\epsilon+U) \right )$ representing the inverse tunneling lifetime due to coupling to the contacts. Here, $f_{\alpha}(\epsilon)=f \left ( \frac{\epsilon-\mu_{\alpha}}{k_BT_{\alpha}} \right )$ 
refers to the Fermi-Dirac distribution of either contact held at an electrochemical potential $\mu_{\alpha}$ and at a temperature $T_{\alpha}$. The many body exchange field may be interpreted as a magnetic field $\vec{B}_{\alpha}=p_{\alpha} \frac{\Gamma_{\alpha} \hat{m}_{\alpha}}{\pi \hbar} \int^{\prime}{dE \left ( \frac{f(E)}{E-\epsilon-U}+\frac{1-f(E)}{E-\epsilon} \right ) }$, with the prime in the integral denoting the Cauchy principal value. The expression for the terminal charge current $J^q_{\alpha}$ is given by:
\begin{eqnarray}
J^q_{\alpha} &=& \frac{2q}{\hbar} \Gamma_{\alpha} [ f_{\alpha}(\epsilon) \rho_{00}  \nonumber\\
&+& \frac{1-f_{\alpha}(\epsilon) +f_{\alpha}(\epsilon+U)}{2} (\rho_{11} + \rho_{22}) \nonumber\\
&-& (1-f_{\alpha}(\epsilon +U) ) \rho_{33}  \nonumber\\
&-& p_{\alpha}[(1-f_{\alpha}(\epsilon) + f_{\alpha}(\epsilon +U)]\hat{m}_{\alpha} \cdot \vec{S} ]. 
\label{eq:curr_def}
\end{eqnarray}
In the above equation, the current depends on the dot occupation probabilities given in terms of the diagonal terms of the density matrix $\rho_{ii}$ and also the dot spin vector $\vec{S}$. In the absence of spin flip processes, one may deduce the expression for terminal spin currents via a simple continuity equation based on Eq. (\ref{eq:spin_dyn}) for the spin accumulation in the quantum dot as $\frac{2q}{\hbar}\frac{d\vec{S}}{dt}=\vec{J}^s_L+\vec{J}^s_R$, where $\vec{J}^s_{L(R)}$ is the terminal spin current with its three components representing transport of $\hat{x}$, $\hat{y}$, and $\hat{z}$ polarized spins along the direction of electrical current \cite{Ralph_Stiles_2007}. One may then write an expression for the terminal spin currents as \cite{Koenig_3}:
\begin{eqnarray}
\vec{J}^s_{\alpha} &=&  J^q_{\alpha}p_{\alpha}\hat{m}_{\alpha} - \frac{2q}{\hbar} \left ( \frac{\vec{S}-p^2_{\alpha}(\hat{m}_{\alpha} \cdot \vec{S})\hat{m}_{\alpha}}{\tau_{r,\alpha}} \right ) -  \frac{2q}{\hbar} \vec{S} \times \vec{B}_{\alpha} \nonumber\\
&=&\vec{J}^s_{\alpha,\hat{m}_{\alpha}} + \frac{2q}{\hbar} \left (\frac{d \vec{S}}{dt} \right )_{\alpha,rel} + \frac{2q}{\hbar} \left (\frac{d \vec{S}}{dt} \right )_{\alpha,prec},
\label{eq:def_spin_curr}
\end{eqnarray}
with $\vec{J}^s_{\alpha,\hat{m}_{\alpha}}$ representing the component due to injection, which is in the direction of magnetization of the contact,  $\frac{2q}{\hbar} \left (\frac{d \vec{S}}{dt} \right )_{\alpha,rel}$ and  $\frac{2q}{\hbar} \left (\frac{d \vec{S}}{dt} \right )_{\alpha,prec}$ representing the angular momentum transfer rate, in units of charge current, due to relaxation and precession, respectively. The first term has a straightforward interpretation simply as being the spin current carried by a spin polarized charge current. The other terms represent angular momentum transfer rates associated with either contact. Specifically, the precession term that arises from a field-like spin torque $\tau_{\alpha}=\left ( \frac{d\vec{S}}{dt} \right )_{\alpha,prec}=\vec{S} \times \vec{B}_{\alpha}$ represents an angular momentum transfer transverse to the magnetization of the contact and to the spin in the dot. This term, although it has a qualitatively different flavor in comparison to the first, it may still be viewed as a spin current \cite{Koenig_3}. Therefore, in this paper, when we talk of spin currents, it is the net terminal spin current given in Eq.~(\ref{eq:def_spin_curr}) that is being considered. 
\\ \indent The relative contribution of spin injection, damping and precession terms that are described by the first, second and the third term in 
Eq. (\ref{eq:spin_dyn}) may be tuned relative to each other via the application of a gate and bias voltage. We therefore focus on the spin blockade regime in which a sizeable spin accumulation may be achieved, and where the relaxation and injection terms are vanishingly small in comparison to the precession term. A sample transport energy configuration of the considered set up is depicted in Fig.~\ref{fig:sp_fig1}(d) where spin accumulation may be induced via spin filtering. The accumulation is usually directed anti-parallel to the spin polarization of the collector contact.
\subsection{Transport set up} We consider transport across the set up shown in Fig.~\ref{fig:sp_fig1}(a). The relative angle between the two contacts is taken as $\theta =\pi/2$, with the left contact being polarized in the $\hat{x}$ direction and the right contact being polarized in the $\hat{z}$ direction. Indeed such a configuration has been experimentally realized in the context of spin torque oscillators \cite{Ralph_2} using a magnetic free layer as the channel. We consider two cases: I) Symmetric case: the polarizations of the two contacts are identical, $p_L=p_R$; II) Asymmetric case: the polarizations of the two contacts are different, $p_L \neq p_R$, 
making one contact of larger polarization in comparison to the other. The asymmetry in the degree of polarization has a profound consequence when a pure temperature gradient is applied. As we will show in the upcoming analysis, due to this asymmetry, a minor imbalance in the tunneling rates between the addition and removal process in the set up created by a pure temperature gradient is enough to induce a non-equilibrium spin accumulation due to spin blockade and hence trigger a spin precession. We take $p_L=p_R=1$ for the symmetric case, and $p_L=1,p_R=0.2$ for the asymmetric case. For our transport set up, we take the contact couplings to be $\hbar \Gamma =0.01 meV$; the Coulomb interaction parameter is $U=40k_BT_L$. When no temperature gradient is applied, we choose $T_R=T_L=0.7K$. In the case of thermoelectric transport we have $T_R=0.7K$ and $T_L=0.9K$. 
\\ \indent The important spin transport effects to be discussed in this paper focus on the regime of blockade and specifically around zero-bias where charge currents are vanishingly small. There is hence a possibility of higher order transport processes such as cotunneling and Kondo effect influencing the physics of transport in this regime. For example, it has been shown in the case of a collinear quantum dot spin valve set up that spin-flip cotunneling processes \cite{Schoen} may significantly influence the spin accumulation as well as the overall conductivity close to zero bias. This happens specifically when the tunnel coupling energy becomes of the order of the ambient temperature or higher ($\hbar \Gamma \geq k_BT$) although the ambient temperature may be well above the Kondo temperature. Furthermore, in the non-collinear set up that we consider here, the fourth order expansion will involve a larger class of two-electron tunneling mechanisms resulting from the coherence terms of the density matrix \cite{Koller}. Therefore, the magnitude of the tunnel coupling energy relative to the ambient temperature must satisfy $\hbar \Gamma < k_BT$ ($9 \hbar \Gamma \approx k_BT_L$  in our case) so that the predictions made here out of the second order theory may remain valid atleast in the conducting region. This also ensures that the ambient temperature is well above the Kondo temperature and hence the influence of Kondo physics on the zero bias transport is also absent. The results presented here are certainly a reasonable approximation close to the boundary of the Coulomb blockade region, while deep inside the blockade, cotunneling might (or might not) alter the findings which could be the subject of a future study.
\begin{figure}
	\centering
		\includegraphics[width=3.6in,height=3.3in]{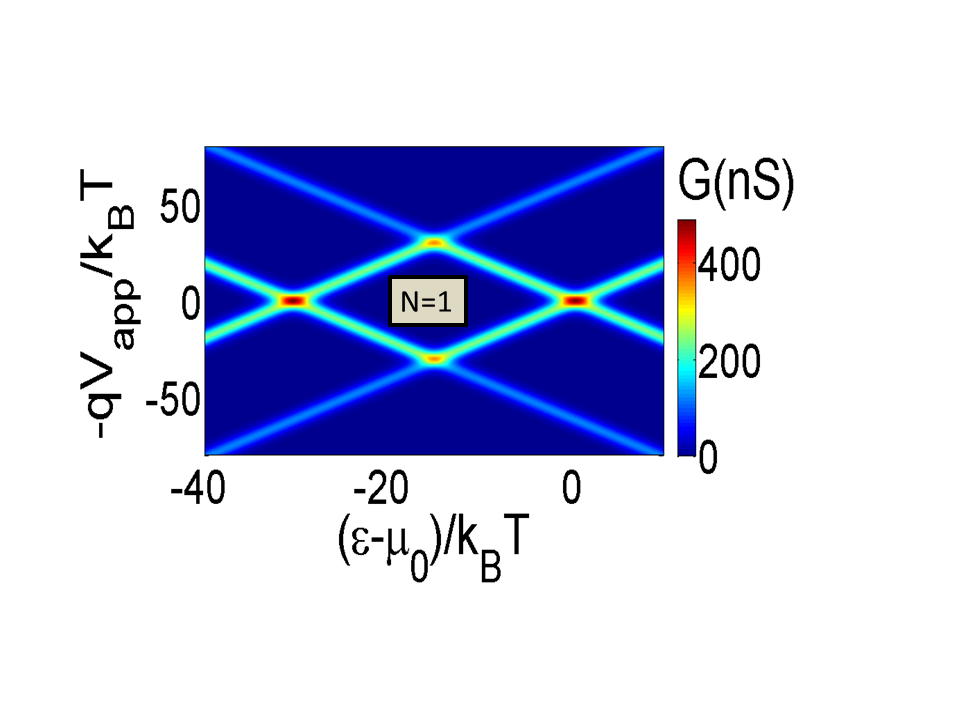}
		\caption{(Color online). Differential conductance $G=\frac{dJ_q}{dV_{app}}$ versus gate and bias voltages showing the $N=1$ sector for the unpolarized case. Here the diamond edges mark the entry of a conducting energy level of the dot within the transport window. We choose $T_R=T_L=0.7K$, $\hbar\Gamma =0.01 meV$ and $U=40k_BT_L$.} 
	\label{fig:sp_fig_extra}
\end{figure}
\\ \indent 
\begin{figure}
	\centering
		\includegraphics[width=3.6in,height=4.5in]{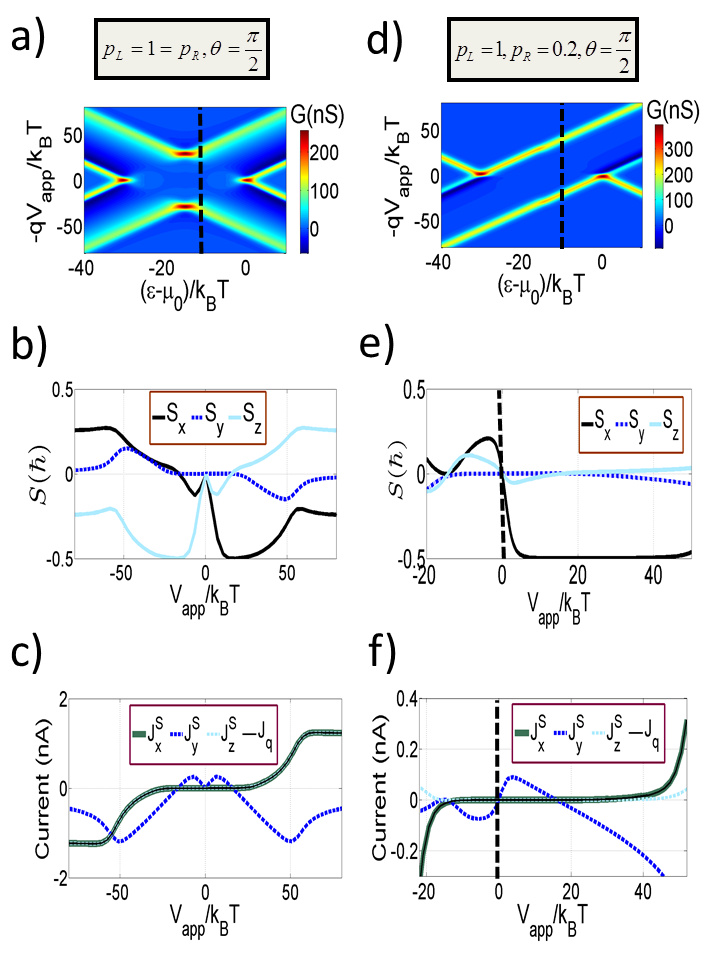}
		\caption{(Color online) Spin accumulation and currents. a) Stability plot depicting the $N=1$ sector for the case of symmetric polarization $p_L=p_R=1$, and $\theta=\pi/2$. Non-equilibrium spin 
transport across the black cut is considered in Figs. b), c).
b) Spin accumulation vs applied bias indicating an $S_x=-1/2$ blocking state in the forward bias direction and an $S_z=-1/2$ blocking state in the reverse bias direction (see text). 
c) Resulting charge and spin currents depicting pronounced $\hat{y}$-polarized spin currents $J^S_y$ in the region of Coulomb blockade. d) Stability plot in the case of the asymmetric polarization $p_L=1$, $p_R=0.2$ case. In this 
case the spin blockade and hence spin accumulation only occurs along the forward bias direction. e) and f) Resulting spin accumulation and currents according to the dashed line in Fig. d). Coulomb blockaded regions at finite bias voltages feature sizeable transversely polarized spin currents with vanishingly small charge and in-plane spin currents. Remaining parameters are the same as in Fig.~\ref{fig:sp_fig_extra}. } 
	\label{fig:sp_fig2}
\end{figure}
\section{Results} 
\subsection{Spin blockade effects}
The spin blockade mechanism relevant to our set up is critical in understanding the occurrence of the non-equilibrium spin accumulation. We hence first set out to illustrate how spin blockade may be identified in the aforementioned cases using a stability plot, i.e., a plot of the differential conductance $G=\frac{dJ_q}{dV_{app}}$ in a bias voltage and gate voltage plane. Here, the effect of the application of a gate field has been encapsulated as an effective detuning $\frac{\epsilon-\mu_0}{k_BT}$ of the energy level $\epsilon$ with respect to the equilibrium electrochemical potential $\mu_0$. Here, $\mu_0$ has been arbitrarily set at the transition energy between the $0$ particle and the $1$ particle configurations. At a finite applied bias of $V_{app}$, assuming equal capacitive couplings of the dot with the two contacts, the contact electrochemical potentials are given by $\mu_L=\mu_0+qV_{app}/2$, and $\mu_R=\mu_0-qV_{app}/2$. In general, there could be an asymmetric voltage drop due to unequal capacitive couplings leading to a distortion in the Coulomb blockade region in the stability plot. The stability plot for an unpolarized set up is shown in Fig.~\ref{fig:sp_fig_extra}. Upon increasing the bias beyond the Coulomb blockade regime, one reaches the diamond {\it{edges}}, signaling the fact that the conducting energy level of the dot has now entered the transport window, and hence transitions between the $0$ particle and $1$ particle configurations are energetically allowed. However, in comparison with the unpolarized case shown in Fig.~\ref{fig:sp_fig_extra}, those diamond edges are clearly absent in Fig.~\ref{fig:sp_fig2}(a), and present along only one bias direction in Fig.~\ref{fig:sp_fig2}(d), indicating spin-blockade. Along the black cut shown in Fig.~\ref{fig:sp_fig2}(a), which corresponds to $\epsilon-\mu_0=-12k_BT_L$, for example, the conducting transition is expected to occur at an applied bias of $V_{app}=V_T=\pm 24 k_BT_L/q$.  We will consider spin dependent transport along this black cut in the analysis to follow, by first elucidating the mechanism of spin blockade in our considered set up.
\\ \indent The plots of relevant transport properties as a function of applied bias for the symmetric and asymmetric case are shown in the left and right panels of Fig.~\ref{fig:sp_fig2} respectively. The spin blockade regime crucial to this work is qualitatively different from the ones observed in the double quantum dot structure, and occurs based on the following mechanism. Consider transport along the black cut in the stability diagrams in Figs.~\ref{fig:sp_fig2}(a) and ~\ref{fig:sp_fig2}(d). Along the forward bias direction, in our convention, the right contact is the injector and the left contact is the collector. In both the symmetric and the asymmetric case, spins injected from the right contact are in varying degrees $+\hat{z}$ polarized, while the left contact that acts as the collector is fully polarized along the $+\hat{x}$ direction, and acts as a spin filter accepting only $S_x=+1/2 \hbar$ electrons. By noting that $\mid S_z=\pm 1/2 \rangle =\frac{1}{\sqrt{2}}(\mid S_x=+1/2 \rangle \pm \mid S_x=-1/2 \rangle )$, the spin filtering at the acceptor leaves behind an accumulation of $S_x=-1/2 \hbar$ spins in the dot. This results in a transport blockade as the energetics forbid the blocked electrons to tunnel back to the right contact. In the reverse bias situation, excess spins along the $-\hat{z}$ direction accumulate to produce a similar blockading effect for the symmetric case. The bias range of the blockade is affected by the polarization of the collector contact.  The effectiveness of the reverse bias blockading effect for the asymmetric case, therefore, is considerably diminished since the right (collector) contact is partially polarized. The spin blockade regime can be observed in Figs.~\ref{fig:sp_fig2}(b) and ~\ref{fig:sp_fig2}(e) in the excess accumulation of spins along the $-\hat{x}$ or $-\hat{z}$ directions, and the suppression of charge and in-plane spin currents in the post-threshold regions of Figs.~\ref{fig:sp_fig2}(c) and ~\ref{fig:sp_fig2}(f). In the symmetric case, this occurs along both bias directions, and in the asymmetric case only along the forward bias direction. 
\\ \indent The polarization of the injecting contact determines the amplitude of the exchange field $\vec{B}_{\alpha}$ associated with it. In turn, the amplitude of the exchange field affects the effectiveness of the torque like term  $-\frac{2q}{\hbar} \vec{S} \times \vec{B}_{\alpha}$ in Eq. (\ref{eq:spin_dyn}) which induces a precession of the accumulated spin in the dot, and hence can partially remove the spin blockade. Because in the asymmetric case the polarization of the injector is smaller than in the symmetric case, the onset of spin blockade and its persistence are more pronounced in this situation as seen by comparing Fig.~\ref{fig:sp_fig2}(b) and Fig.~\ref{fig:sp_fig2}(e).  
\subsection{Spin precession and associated spin currents} 
Due to the prevailing blockade conditions, spins injected from either contact are subject to precessional dynamics on a time scale amounting to the tunneling lifetime. In the steady state, the precessing spin eventually aligns with the net effective magnetic field. In the bias region $0 \leq V \leq V_T$ it can be shown in steady state that, $S_y(V)=0$, and $S_x(V)/S_z(V)=B_L(V)/B_R(V)$. The effective spin accumulation is directed along the effective exchange field direction given by $\vec{B}_{eff}=B_L \hat{m}_L+ B_R \hat{m}_R$. While the steady state solution simply points to the spins being aligned with the effective field such that $\vec{S} \times \vec{B}_{eff}=0$, the field-like spin torques associated with each contact $\tau_{\alpha}=\vec{S} \times \vec{B_{\alpha}}$ do not vanish. As a result, the angular momentum transfer rate and hence spin currents associated with either contact is finite and given as $\vec{J}^s =(\vec{J}^s_L-\vec{J}^s_R)/2$. The associated transversely polarized terminal spin currents are depicted by the $\hat{y}$ component of the spin current tensor (shown dashed blue) in Fig.~\ref{fig:sp_fig2}(c) and (f), and the charge currents and in-plane spin currents in the whole blockade region are effectively negligible. 
\begin{figure}
	\centering
		\includegraphics[width=3.6in,height=3.1in]{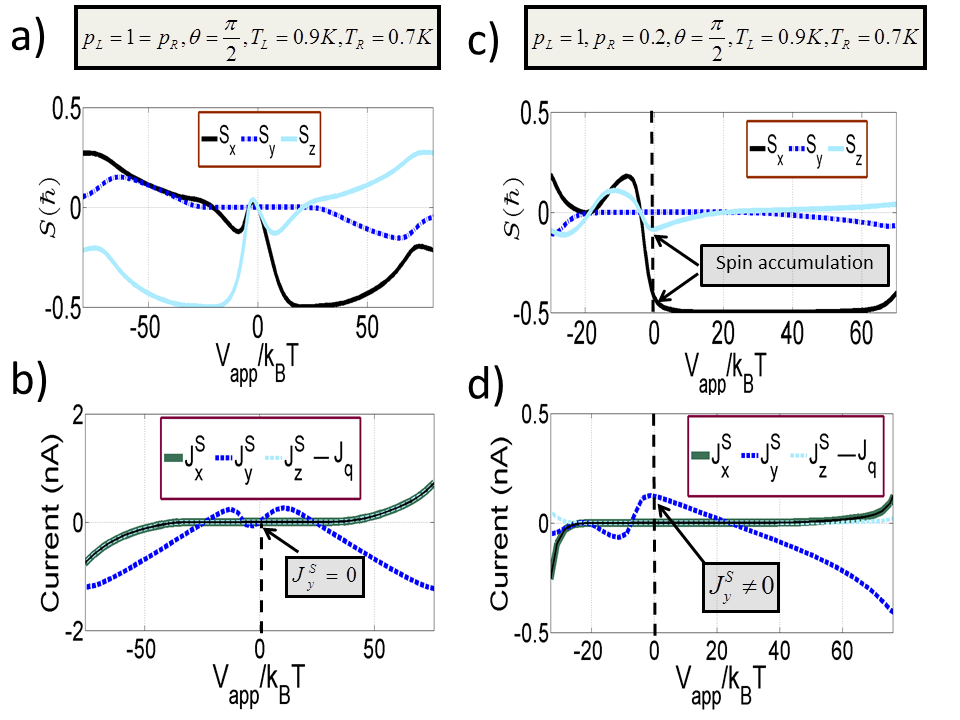}
		\caption{(Color online) Effect of temperature gradient. a) In the case of symmetric polarization, 
for a small temperature gradient, the zero bias spin accumulation is always absent. Thus, spin precession is absent at zero bias, resulting in a b) zero $\hat{y}$ polarized spin current $(J^s_y(V_{app}=0)=0)$. 
c) In the asymmetric case, however, zero bias spin accumulation occurs and the resulting spin precession causes a d) non-zero transverse spin current $(J^s_y(V_{app}=0)\neq 0)$ (blue dashes). The in-plane spin currents and charge currents 
are vanishingly small in this region.}
	\label{fig:sp_fig3}
\end{figure}
\subsection{Effect of a temperature gradient} 
\begin{figure}
	\centering
		\includegraphics[width=3.2in,height=2.9in]{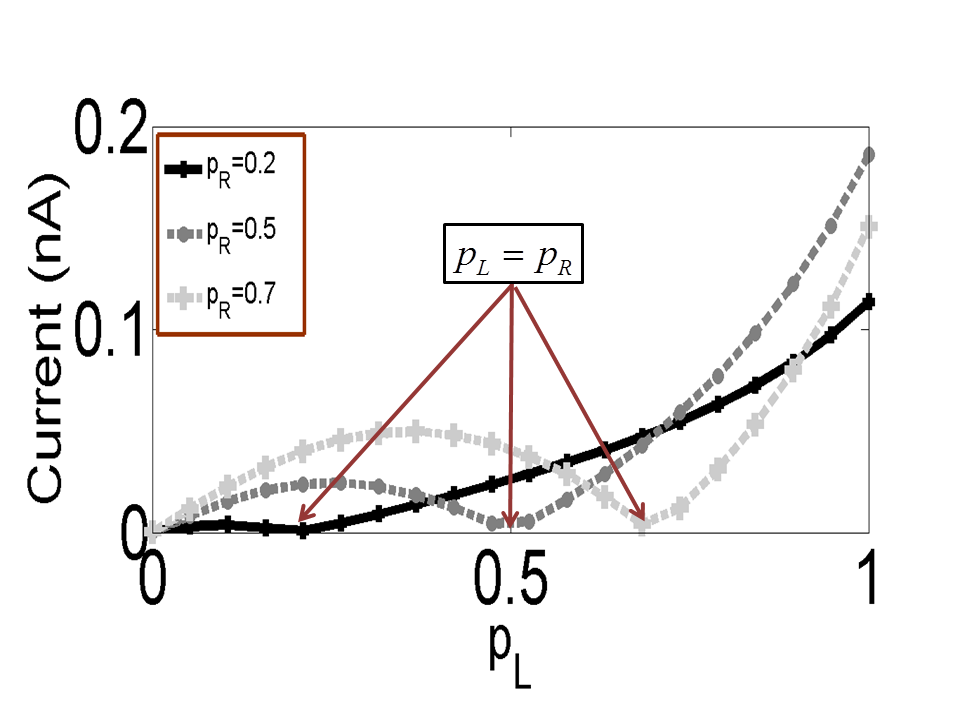}
		\caption{(Color online) Dependence of the zero bias spin current magnitude on the relative polarization between the contacts. The spin current vanishes when $p_L=p_R$ and also when either contact is unpolarized and varies quadratically in between. For $p_L>p_R$ we notice a quasi-linear variation in the spin current magnitude. The noticed trends here indicate the possibility of realizing a sizeable spin current magnitude for a wide range of realistic polarization magnitudes for the two contacts.} 
	\label{fig:sp_fig5}
\end{figure}

In the asymmetric case, as remarked before, the most important consequence of the above discussed spin blockade mechanism is the {\it{zero bias}} non-equilibrium spin accumulation emerging with the application of a temperature gradient. A small temperature gradient (we choose $\Delta T=0.2K$, such that $T_L=0.9K$, and $T_R=0.7K$) in the absence of a bias opens the possibility of charge and spin transport via thermoelectric operation \cite{Basky}. As shown in Fig.~\ref{fig:sp_fig3}(c), the asymmetric situation induces a zero bias spin accumulation due to a small imbalance between the tunneling rates of the left and the right contacts. In contrast, no zero bias spin blockade occurs in the symmetric polarization case shown in Fig.~\ref{fig:sp_fig3}(a).
\begin{figure}
	\centering
		\includegraphics[width=2.9in,height=3.2in]{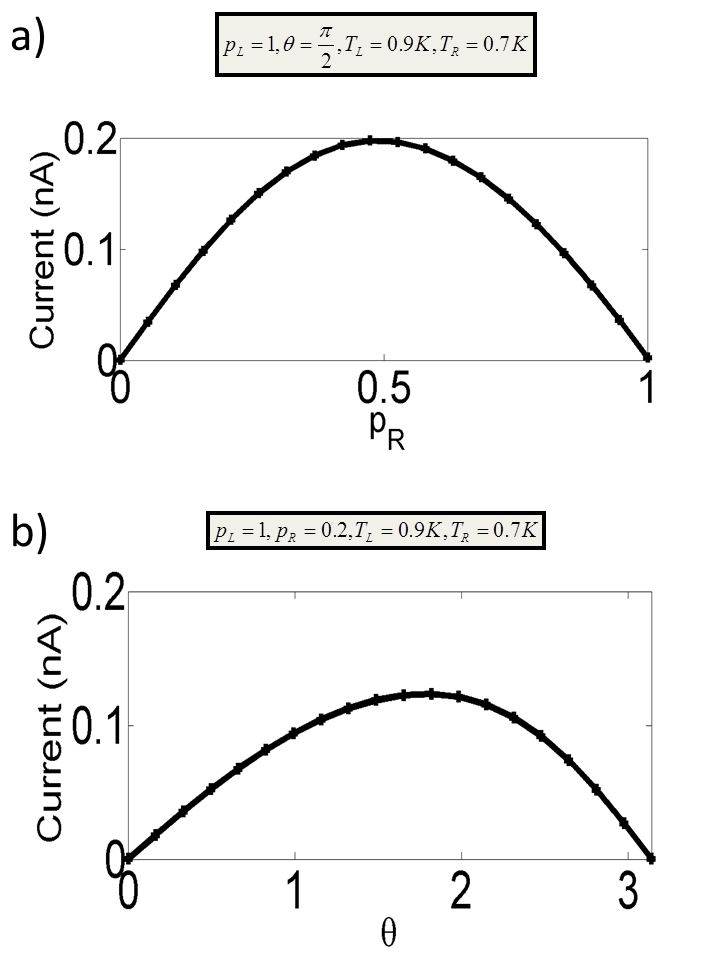}
		\caption{(Color online) Polarization and angular dependence. a) Dependence of the zero bias thermal pure spin current magnitude on the degree of polarization of the right contact. For the unpolarized case and 
the fully polarized cases, the transverse spin current is zero. The pure spin current magnitude peaks at $p_R=0.5$. b) Angular dependence of the magnitude of the pure spin current with $p_R=0.2$. The asymmetry is due to the fact 
that the situation $p_R=0.2$ corresponds to the majority up-spin case.} 
	\label{fig:sp_fig4}
\end{figure}
The accumulation results in a zero bias spin torque $\tau_{\alpha}=\vec{S} \times \vec{B}_{\alpha}$ at either contact and hence in an associated $\hat{y}$-polarized pure spin current as shown in Fig.~\ref{fig:sp_fig3}(d). The spin accumulation is in the plane of magnetization of the two contacts, and the spin precession dynamics due to the third term in Eq. (\ref{eq:spin_dyn}) results in a non-zero rate of transfer of transverse spin angular momentum. 
\\ \indent The result presented here involves the generation of a spin current due to an applied thermal gradient. In order to make a connection with energy conversion, one has to typically quantify the efficiency of this process. The process efficiency depends on the detection method and utilization of the spin current. For example, if this spin current was detected via electrical means, the efficiency of this process would depend on the power drawn in the circuit \cite{Basky}. Typically, in the case of charge thermoelectric effects, the maximum efficiency in the linear response regime is related to a dimensionless metric called the figure of merit $zT$, which is defined as $zT=\frac{S^2 \sigma T}{\kappa}$, where $S$ is the Seebeck coefficient, $\sigma$ is the electrical conductivity, $\kappa$ is the thermal conductivity and $T$ is the ambient temperature. In the collinear polarization case, a few recent works \cite{DiVentra_spin,Barnas_1} have defined a similar spin figure of merit $Z_sT=\frac{S^2_s \sigma_s T}{\kappa}$, where $S_s$ is the spin Seebeck coefficient and $\sigma_s$ is the spin dependent conductivity. The premise of defining a spin dependent figure of merit was motivated by the linear response expansion of the charge, in-plane spin currents and heat currents. In principle, one could extend this for our non-collienar case by a linear response expansion that includes the charge current, the in-plane spin current and the transversely polarized spin currents via a four-component voltage drop \cite{Brataas_1} with an Onsager matrix \cite{Brataas_1,Bauer_2011} that couples with the heat current. However, it is left to a more rigorous analysis to assess the validity as well as the merit in defining performance metrics such as $Z_sT$ for this case.
\\ \indent We now analyze the effect of varying the lead polarizations and angles. Keeping the left contact fully polarized ($p_L=1$), we plot the polarization and angle dependence of the magnitude of this zero-bias transverse spin current in Fig.~\ref{fig:sp_fig4}. It is seen from Fig.~\ref{fig:sp_fig4}(a) that the spin current magnitude 
is zero for the unpolarized and the fully symmetric case and maximizes at $p_R=0.5$. The angular dependence of this spin current magnitude between $\theta=0$ and $\theta=\pi$ is shown in Fig.~\ref{fig:sp_fig4}(b) for $p_R=0.2$. The noted asymmetry simply arises from the fact that the majority spins are along the $S_z=+1/2$ direction. Having either contact fully polarized is still an idealization and was used in this paper in order to elucidate the non-trivial physics that was to be conveyed. Real ferromangetic contact polarizations in the best case approach $30-40$ percent. We therefore study the effect of varying both contact polarizations in Fig.~\ref{fig:sp_fig5} keeping $\theta=\pi/2$ so that a realistic range of contact polarization magnitudes may be assessed. Here, the three curves depict the variation of the zero bias spin current magnitude with the left contact polarization $p_L$ for three representative values of $p_R$. As expected, we note that the spin current magnitude vanishes when either contact is unpolarized and when $p_L=p_R$ and varies quasi-quadratically in between. Furthermore, making $p_L>p_R$ results in a quasi-linear variation of the spin current magnitude. The noticed trends here indicate the possibility of realizing a sizeable spin current for a wide range of realistic polarization magnitudes for the two contacts.
\subsection{Discussion and perspectives} The results presented so far might have important implications. First, the spin accumulation result presented here opens the interesting possibility of spin initialization via a small temperature gradient, in the absence of a bias. Second implication is the occurence of transversely polarized terminal spin currents due to the zero bias field-like spin torque. The relaxation dynamics typically result from a transition from the one electron state into the zero electron state or to the two electron state, both of which are spin zero states. Typical relaxation times in this case are very long and of the order of $10$ $\mu s$. These long-time coherent spin rotations may have important applications with respect to spin manipulation via a gate pulse, such that the spin rotation may be read out once blockade is removed by gating the dot energy level. Furthermore, the precession may be used to probe relaxation times due to other relaxation mechanisms within the quantum dot \cite{SB_2}. 
\\ \indent  Finally, the question of detecting the spin currents discussed here has numerous subtleties. It has been theoretically established \cite{Brataas_1,Brataas_2,Brataas_3} and experimentally demonstrated \cite{Ando_2011} that precessing spins in a free magnetic thin layer that is coupled to pinned ferromagnetic or normal metallic contacts can result in the volume generation of spin currents. The ferromagnetically pinned contact often acts as a spin sink that will absorb the transversely polarized angular momentum flow. However, due to the conservation of angular momentum, a back acting torque will induce a perturbation in the precessing spins. Such a perturbation may be indirectly detected via the broadening of the ferromagnetic resonance lines described in \cite{Brataas_4,Brataas_5}. A more direct method would be to use a free magnetic thin layer within a spin relaxation length in between the collector contact and the quantum dot, and hence detecting the angular momentum transfer via the precession of this layer. Alternatively, the magneto-optic Kerr effect \cite{C_Back} may be used to directly detect the excitation due to this pure spin current. While it is shown in the context of magnetization dynamics that a similar magnetization precession may be related to pure spin currents \cite{Brataas_1,Brataas_2,Brataas_3,Brataas_4,Brataas_5}, progress on understanding the implications of similar phenomena with respect to single spin precession noted here would form an interesting and important extension of this work. 
\vspace{0.02in}
\section{Conclusions} In this paper, we explored spin dependent phenomena in the thermoelectric regime of a non-collinear quantum dot spin valve set up. This work opens the interesting possibility of thermoelectric manipulation of single spins in a quantum dot transport set up. The spin torque and the related spin dynamics discussed here are reminiscent of what is observed in the collective case as a spin torque in the magnetization dynamics of magnetic layers. We showed that when the set up is biased deep into blockade where double occupancy is forbidden, a resulting zero bias thermoelectric spin torque may yield a long time spin precession. The implications of this with respect to single spin manipulation as well as its connection with pure spin currents were discussed. Unlike in the collective case, the spin dynamics inside quantum dot arrays may be thought of as an ensemble of weakly interacting spins. Electrical or thermoelectric control of spin dynamics of individual spins interacting via a quantum dot array may in general open exciting paradigms and possibilities.
\vspace{0.1in}
\\ {\bf{Acknowledgements:}} This work was partly supported by the Deutsche Forschungsgemeinschaft (DFG) under programme SFB 689. BM would like to acknowledge financial support from the IIT Bombay IRCC-SEED grant.
\bibliographystyle{apsrev}	% (uses file "plain.bst")
\bibliography{one_level_spin_pump_bib_2}		% expects file "myrefs.bib"
\end{document}